\documentclass[sigconf]{acmart}

\AtBeginDocument{%
  }

\setcopyright{none}
\copyrightyear{2026}
\acmYear{2026}
\acmDOI{}

\acmConference[ACM-BCB '26]{17th ACM Conference on Bioinformatics, Computational Biology, and Health Informatics}{June 30 -- July 3, 2026}{Calabria, Italy}
\acmBooktitle{ACM-BCB '26}
\acmISBN{}

\usepackage{microtype}
\usepackage{amsmath,amsfonts}
\usepackage{graphicx}
\usepackage{xcolor}
\usepackage{cleveref}
\usepackage{multirow}
\usepackage{enumitem}
\usepackage{float}

\definecolor{detectable}{HTML}{2C5F82}
\definecolor{notdetectable}{HTML}{74AECB}

\title{\textsc{SpliceBind}: Isoform-Aware Prediction of Binding Pocket Druggability}

\author{Bryan Cheng}
\affiliation{%
  \institution{Independent Researcher}
  \country{USA}
}

\author{Austin Jin}
\affiliation{%
  \institution{Independent Researcher}
  \country{USA}
}

\author{Joshua Chang}
\affiliation{%
  \institution{Independent Researcher}
  \country{USA}
}

\begin{abstract}
Splice-mediated drug resistance occurs in up to 40\% of patients on targeted kinase inhibitors, yet state-of-the-art druggability tools operate on single structures and cannot compare across isoforms.
We introduce \textsc{SpliceBind}, a graph neural network framework for isoform-aware druggability prediction.
Beyond improving prediction accuracy (AUROC $0.703$ vs.\ P2Rank $0.634$, $p = 0.026$), we address a more fundamental question: when do structural methods succeed, and when must they fail?
Systematic analysis of six clinically validated variants spanning five mechanism classes reveals a two-tier resistance taxonomy.
Domain deletions (AR-V7, $\Delta = -18.39$) and pocket disruptions produce structurally detectable changes, while allosteric mechanisms (BRAF-p61) remain fundamentally invisible to any pocket-centric approach---a boundary no algorithmic improvement can cross.
Notably, learned embeddings capture affinity-based resistance missed by geometry alone (ALK-L1196M: $\Delta_{\text{SB}} = -0.228$ vs.\ $\Delta_{\text{P2Rank}} = -0.95$), partially bridging the structural--biochemical gap.
On 229 kinase pockets spanning 25 families, \textsc{SpliceBind} achieves AUROC $0.703$ ($p = 0.026$ vs.\ P2Rank) with robust generalization to held-out families (AUROC $0.761$).
This taxonomy transforms clinical workflows: upon discovering a splice variant, clinicians can immediately determine whether computational triage suffices or biochemical validation is required---reducing time from variant discovery to therapeutic decision.
\end{abstract}



\begin{document}
\maketitle

\section{Introduction}
\label{sec:intro}

Alternative splicing generates protein isoforms that increasingly drive therapeutic resistance in cancer.
The androgen receptor variant AR-V7 confers enzalutamide resistance in prostate cancer~\cite{antonarakis2014ar}, BRAF-p61 drives vemurafenib resistance in melanoma via RAS-independent dimerization~\cite{poulikakos2011raf}, and MET-exon14skip remains druggable by capmatinib despite being oncogenic~\cite{paik2020response}.
Aberrant splicing is increasingly recognized as a driver of therapeutic resistance~\cite{dvinge2016rna}, yet the field lacks tools to predict whether a given splice variant will retain or lose drug sensitivity.

The clinical scale of this problem is substantial.
AR-V7 is detected in 20--40\% of patients who progress on androgen-targeting therapies~\cite{antonarakis2014ar,scher2016ar}, BRAF splice variants account for roughly a third of vemurafenib-resistant melanomas~\cite{poulikakos2011raf}, and pan-cancer analyses have identified thousands of tumor-specific splicing events across 33 tumor types~\cite{kahles2018comprehensive}.
Yet validating the functional consequences of each variant currently requires months of biochemical and cellular assays.
Targeted kinase inhibitors represent one of the most validated drug classes~\cite{hopkins2002druggable}, and computational pre-screening of splice variants by their structural impact on drug binding would substantially accelerate clinical decision-making.

Existing druggability tools (P2Rank~\cite{krivak2018p2rank}, fpocket~\cite{le2009fpocket}, SiteMap~\cite{halgren2009identifying}) predict binding pockets in individual structures but were not designed for isoform-level comparison.
They cannot answer the key clinical question: \emph{does this splice variant's binding pocket differ from the canonical form in a way that affects drug response?}
Protein language models like ESM-2~\cite{lin2023evolutionary} capture evolutionary context that could complement geometric features, but their integration into splice variant analysis is unexplored.

We present \textsc{SpliceBind}, a graph neural network (GNN) framework that predicts pocket druggability using structural features augmented with ESM-2 embeddings.
Our contributions are threefold:
\begin{enumerate}[nosep,leftmargin=*]
  \item \textsc{SpliceBind}, an isoform-aware druggability prediction framework, combining pocket graph neural networks with protein language model embeddings to compare canonical and variant binding sites;
  \item A two-tier resistance taxonomy: structurally detectable vs.\ not, comprising five mechanism subtypes (domain deletion, pocket disruption, preserved druggability, allosteric, affinity-based) derived from systematic analysis of clinically relevant splice variants;
  \item An empirical boundary analysis demonstrating which mechanism classes are structurally detectable and which require complementary biochemical validation.
\end{enumerate}
Through six case studies spanning five distinct mechanism classes, we find that domain deletions and pocket disruptions are robustly detectable by structural methods, affinity-based mechanisms are partially captured by learned features, and allosteric mechanisms require biochemical approaches.
This taxonomy guides the selection of validation strategies when new splice variants are discovered.

\section{Related Work}
\label{sec:related}

\paragraph{Binding site detection.}
Geometry-based methods such as fpocket~\cite{le2009fpocket} use Voronoi tessellation to identify cavities.
SiteMap~\cite{halgren2009identifying} scores pockets via van~der~Waals and electrostatic fields; P2Rank~\cite{krivak2018p2rank} applies random forests to local chemical features; DoGSiteScorer~\cite{volkamer2012dogsitescorer} combines grid-based detection with druggability scoring.
Learned approaches such as DeepSite~\cite{jimenez2017deepsite} and Kalasanty~\cite{stepniewska2020improving} predict pockets from voxelized 3D grids using CNNs, achieving competitive accuracy on single-structure benchmarks.
However, none of these methods---geometry-based or learned---support isoform-level comparison or incorporate evolutionary sequence context.

\paragraph{Protein language models.}
ESM-2~\cite{lin2023evolutionary,rives2021biological} learns contextualized per-residue embeddings enabling structure prediction and capturing functional information beyond geometry.
ProtTrans~\cite{elnaggar2021prottrans} demonstrated effective transfer to downstream protein tasks.
Integration of such embeddings into druggability prediction---specifically splice variant analysis---has not been explored.

\paragraph{GNNs for molecular tasks.}
SchNet~\cite{schutt2017schnet} introduced continuous-filter convolutions for molecular graphs; GVP~\cite{jing2021equivariant} extended equivariant networks to macromolecular structures; EdgeConv~\cite{wang2019dynamic} enables dynamic local graph neighborhoods for point clouds and protein surfaces.
MaSIF~\cite{gainza2020deciphering} and EquiBind~\cite{stark2022equibind} applied geometric deep learning to protein interaction and docking tasks; DeepFRI~\cite{gligorijevi2021structure} demonstrated GNN-based protein function prediction from contact maps.
Our work applies EdgeConv to pocket graphs augmented with protein language model features---a combination not previously used for druggability prediction.

\paragraph{Variant impact prediction.}
Sequence-level tools such as SIFT~\cite{ng2003sift} and PolyPhen-2~\cite{adzhubei2010method} predict whether amino acid substitutions are deleterious based on evolutionary conservation and physicochemical properties.
At the structural level, FoldX~\cite{schymkowitz2005foldx} estimates stability changes ($\Delta\Delta G$) from point mutations using empirical force fields, and AlphaMissense~\cite{cheng2023accurate} leverages AlphaFold-derived structural context for proteome-wide pathogenicity classification.
All of these tools are designed for single-residue substitutions: they cannot model the domain-level rearrangements---exon deletions, domain truncations, reading frame shifts---characteristic of splice variants.
\textsc{SpliceBind} addresses this gap by comparing full pocket graphs between canonical and variant structures rather than evaluating individual residue changes.

\paragraph{Drug resistance prediction.}
Computational resistance prediction typically operates at the mutation level using docking~\cite{davis2011comprehensive}, molecular dynamics, or binding affinity models~\cite{hata2016tumor}.
Splicing resources such as VastDB~\cite{tapial2017atlas} catalog events but do not predict functional consequences for drug binding.
Binding affinity prediction methods such as OnionNet~\cite{zheng2019onionnet} and KDEEP~\cite{jimenez2018kdeep} predict protein--ligand $\Delta G$ from co-crystal structures using deep learning, but require known ligand poses as input---unavailable for novel splice variants without prior structural characterization.
These approaches are complementary to \textsc{SpliceBind}'s ligand-free pocket analysis: our framework identifies \emph{whether} a pocket is druggable, while affinity predictors estimate \emph{how tightly} a specific ligand binds.
Methods designed for splice variant--level resistance prediction are absent from the literature.

\section{Methods}
\label{sec:methods}

\begin{figure*}[t]
\centering
\includegraphics[width=\linewidth]{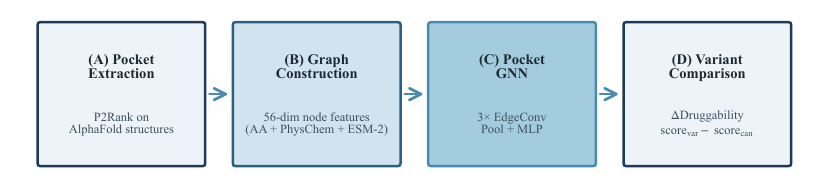}
\caption{
\textbf{\textsc{SpliceBind} architecture.}
\textbf{(A)}~Pocket extraction: P2Rank identifies binding pockets in AlphaFold-predicted kinase structures; ESM-2 extracts per-residue embeddings from the sequence.
\textbf{(B)}~Graph construction: pocket residues become nodes (56-dimensional features) connected by edges within 6\,\AA.
\textbf{(C)}~SpliceBind: three EdgeConv layers followed by global pooling and MLP predict druggability probability.
\textbf{(D)}~Splice variant comparison: $\Delta$Druggability between canonical and variant structures quantifies resistance risk.
}
\Description{Four-panel pipeline diagram showing pocket extraction from AlphaFold structures, graph construction with 56-dimensional node features, SpliceBind architecture with three EdgeConv layers, and splice variant druggability comparison.}
\label{fig:overview}
\end{figure*}

\textsc{SpliceBind} represents each binding pocket as a graph, combines structural and sequence-derived features, and predicts a druggability score that can be compared across splice variants (\Cref{fig:overview}).

\subsection{Dataset Construction}
\label{sec:dataset}

We curated a benchmark of kinase binding pockets with experimentally validated drug binding labels (\Cref{tab:dataset_stats}).
One hundred fifty-three kinase structures from the AlphaFold Protein Structure Database (v6)~\cite{jumper2021highly,varadi2022alphafold} spanning 48 genes across 25 families were processed by P2Rank~\cite{krivak2018p2rank}, yielding 229 binding pockets.
We aggregated drug-kinase binding affinities ($K_d$/IC$_{50}$) from BindingDB~\cite{liu2007bindingdb}, ChEMBL~\cite{gaulton2012chembl}, and the Davis kinase panel~\cite{davis2011comprehensive} to assign experimental labels:
we labeled pockets overlapping known ATP binding sites with a potent binder ($K_d < 100$\,nM) as positive ($n=159$, 69.4\%),
and pockets with no binding site overlap or only weak binders ($K_d > 10$\,\textmu M) as negative ($n=70$, 30.6\%).
Crucially, we use experimental binding data---not P2Rank scores---as ground truth, avoiding circularity in baseline comparisons.
Defining negative samples for druggability prediction is inherently challenging: the absence of a reported binder does not guarantee a pocket is undruggable---it may simply be understudied.
We adopt a conservative labeling strategy, requiring both lack of binding site overlap \emph{and} absence of potent binders ($K_d > 10$\,\textmu M) to assign a negative label.
Pockets with intermediate affinity ($100\,\text{nM} < K_d < 10\,\text{\textmu M}$) are excluded ($n = 7$) to avoid ambiguous labels.
The resulting 69:31 positive-to-negative ratio reflects the kinase-focused nature of our dataset---kinases are among the most heavily drugged protein families, so the majority of well-characterized pockets have known binders.
We address this moderate class imbalance through focal loss with inverse-frequency class weighting (\Cref{sec:training}).

\begin{table}[t]
\centering
\caption{Dataset statistics for 229 kinase binding pockets.}
\label{tab:dataset_stats}

\small
\begin{tabular}{@{}lr@{}}
\toprule
\textbf{Property} & \textbf{Count} \\
\midrule
Total pockets & 229 \\
Positive (experimental binding) & 159 (69.4\%) \\
Negative (no overlap / weak binding) & 70 (30.6\%) \\
Kinase families & 25 \\
Unique genes & 48 \\
Source structures & 153 \\
\bottomrule
\end{tabular}
\end{table}

\subsection{Model Architecture}
\label{sec:architecture}

Each pocket is represented as a graph $G = (V, E)$ where nodes are residues and edges connect C$_\alpha$ atoms within 6.0\,\AA.

\paragraph{Node features (56-dimensional).}
Each residue is described by: (i)~a 20-dimensional one-hot amino acid encoding; (ii)~4 physicochemical properties (hydrophobicity, charge, aromatic indicator, polarity); and (iii)~a 32-dimensional ESM-2 projection.

\paragraph{Why EdgeConv.}
EdgeConv~\cite{wang2019dynamic} dynamically computes edge features from node pair differences, capturing local geometric context without requiring a fixed graph topology.
Unlike message-passing GNNs (e.g., GIN, GAT) that aggregate neighbor features directly, EdgeConv learns edge functions that encode relative spatial and chemical relationships between residue pairs---important for detecting subtle pocket shape changes between canonical and variant structures.
This design is well-suited to pocket graphs, where the number of residues and their spatial arrangement vary substantially across pockets and isoforms.

\paragraph{ESM-2 integration.}
We extract 1280-dimensional per-residue embeddings from the ESM-2 650M-parameter model~\cite{lin2023evolutionary}.
To reduce dimensionality without trainable parameters, we apply binned mean pooling: the 1280-dimensional vector is partitioned into 32 contiguous bins of 40 dimensions, and the mean within each bin yields a 32-dimensional representation.
We chose deterministic binned mean pooling over PCA or a learned projection to avoid introducing trainable parameters at the embedding reduction stage, preserving the pre-trained ESM-2 representation without risk of overfitting on our limited training set.
This fixed reduction retains 82.5\% of pairwise cosine similarity structure, comparable to PCA with 32 components (79.8\%; Appendix~\ref{app:features}).

\paragraph{Network.}
The \textsc{SpliceBind} consists of three EdgeConv~\cite{wang2019dynamic} layers (256 hidden dimensions), concatenated mean and max global pooling, and a three-layer MLP ($512 \to 256 \to 64 \to 1$) with sigmoid output producing druggability probability $p \in [0,1]$.

\paragraph{Model capacity.}
The SpliceBind has 899,457 trainable parameters trained on ${\sim}180$ samples per fold, yielding a parameter-to-sample ratio of ${\sim}5{,}000{:}1$.
We mitigate overfitting through: (i)~aggressive dropout (0.2 in EdgeConv layers, 0.3 in MLP); (ii)~focal loss with class weighting preventing degenerate majority-class solutions; (iii)~early stopping on validation AUROC with patience 10; and (iv)~family-level GroupKFold ensuring the model cannot memorize family-specific patterns.
Despite the high ratio, each pocket graph averages 18.4 residue nodes, providing substantial within-graph supervision signal---the model sees ${\sim}3{,}300$ node-level feature vectors per fold, not merely 180 scalar labels.

\subsection{Training and Evaluation}
\label{sec:training}

We train with focal loss~\cite{lin2017focal} to handle the moderate class imbalance (69:31) (standard cross-entropy produces degenerate majority-class predictions):
\begin{equation}
  \mathcal{L}_{\text{focal}} = -\alpha_t (1 - p_t)^{\gamma} \log(p_t)
  \label{eq:focal}
\end{equation}
with $\gamma = 2.0$ and $\alpha_t$ inversely proportional to class frequency ($\alpha_{\text{pos}} \approx 0.31$, $\alpha_{\text{neg}} \approx 0.69$).
Optimization uses AdamW (lr $= 10^{-3}$, weight decay $10^{-4}$) with ReduceLROnPlateau (factor 0.5, patience 5), early stopping (patience 10), batch size 32, up to 50 epochs.

Evaluation uses 5-fold GroupKFold cross-validation~\cite{krstajic2014cross,roberts2017cross} grouped by kinase family, repeated across 3 seeds (15 folds total).
No kinase family appears in both training and test sets within a fold, providing a stringent generalization test.

\subsection{Splice Variant Analysis}
\label{sec:splice_analysis}

For splice variant evaluation, we (1)~generate variant structures using ESMFold~\cite{lin2023evolutionary}, (2)~apply P2Rank pocket detection to canonical and variant structures, and (3)~compute $\Delta\text{Druggability} = \text{Score}_{\text{var}} - \text{Score}_{\text{can}}$.
$\Delta$Druggability uses P2Rank pocket scores, which quantify structural druggability and are not bounded to $[0,1]$.
Our framework employs a dual-method design.
\textsc{SpliceBind}'s GNN predicts druggability probability $p \in [0,1]$ (\Cref{sec:architecture}), validating that learned features improve upon structural baselines (\Cref{sec:cv_results}).
The splice variant taxonomy uses P2Rank's upstream pocket scores to assess structural changes (\Cref{sec:splice_results}).
We use P2Rank scores for variant comparison because P2Rank operates without training data, providing a training-independent structural baseline that generalizes to novel variants outside the training distribution.
\textsc{SpliceBind} demonstrates that learned predictions improve upon P2Rank, while P2Rank provides the unbiased structural comparison needed for the taxonomy.
We note that ESMFold is built on ESM-2 internally; variant structure quality may therefore correlate with the ESM-2 features used by SpliceBind.
We mitigate this potential coupling by extracting ESM-2 embeddings from sequences, while ESMFold structures serve only for pocket geometry.
We selected six clinically relevant cases spanning the major known mechanism classes reported in oncology literature: AR-V7, MET-exon14skip, PIK3CD-S, EGFRvIII, BRAF-p61, and ALK-L1196M.
The first five are splice variants; ALK-L1196M is a point mutation included as a boundary test case to probe where pocket-level structural analysis fails for single-residue perturbations.

\subsection{Graph Comparison Methodology}
\label{sec:graph_comparison}

We emphasize that canonical and variant pocket graphs are \emph{not aligned, matched, or superimposed}.
Instead, our framework independently predicts druggability for each structure and compares scalar scores---a key methodological distinction from graph-matching or node-correspondence approaches.

\textbf{Pipeline for variant comparison}:
\begin{enumerate}[nosep,leftmargin=*]
  \item Generate variant protein structure via ESMFold~\cite{lin2023evolutionary} from the spliced sequence
  \item Detect binding pockets in both canonical (AlphaFold) and variant (ESMFold) structures using P2Rank~\cite{krivak2018p2rank}
  \item For each detected pocket, construct a residue graph with nodes (pocket residues) and edges (C$_\alpha$ contacts within 6\,\AA)
  \item Extract per-residue ESM-2 embeddings from the full-length sequences (canonical and variant independently)
  \item Independently run \textsc{SpliceBind} inference: $\text{Score}_{\text{can}} = f_{\theta}(G_{\text{can}})$ and $\text{Score}_{\text{var}} = f_{\theta}(G_{\text{var}})$
  \item Compute $\Delta\text{Druggability} = \text{Score}_{\text{var}} - \text{Score}_{\text{can}}$
\end{enumerate}

A negative $\Delta$ indicates reduced druggability in the variant (resistance-associated), while $\Delta \approx 0$ suggests preserved druggability.
Crucially, graphs $G_{\text{can}}$ and $G_{\text{var}}$ differ in size (number of residues), connectivity (edge patterns), and node features (ESM-2 embeddings reflect sequence changes), precluding node-level alignment or correspondence.
The EdgeConv architecture's permutation invariance and global pooling (mean + max aggregation across all nodes) enable comparison of structurally dissimilar pockets via learned fixed-dimensional druggability embeddings.
Each graph is reduced to a 512-dimensional vector before classification, allowing pockets of arbitrary size and topology to be scored on a common scale.

For the resistance taxonomy (\Cref{sec:splice_results}), we use P2Rank's unbounded pocket scores rather than \textsc{SpliceBind}'s sigmoid probabilities to provide interpretable absolute magnitudes ($\Delta = -18$ vs.\ $\Delta = -0.03$ clearly distinguishes mechanism severity) and training-independent generalization to prospective variants.
\Cref{tab:method_comparison} compares both methods' performance on all six clinical variants.

\section{Results}
\label{sec:results}

We first validate that \textsc{SpliceBind} produces meaningful druggability predictions (\Cref{sec:cv_results}), then present our primary contribution: a taxonomy of splice variant resistance mechanisms and the conditions under which structural prediction succeeds or fails (\Cref{sec:splice_results}).

\subsection{Model Validation}
\label{sec:cv_results}

\begin{figure*}[t]
\centering
\includegraphics[width=\linewidth]{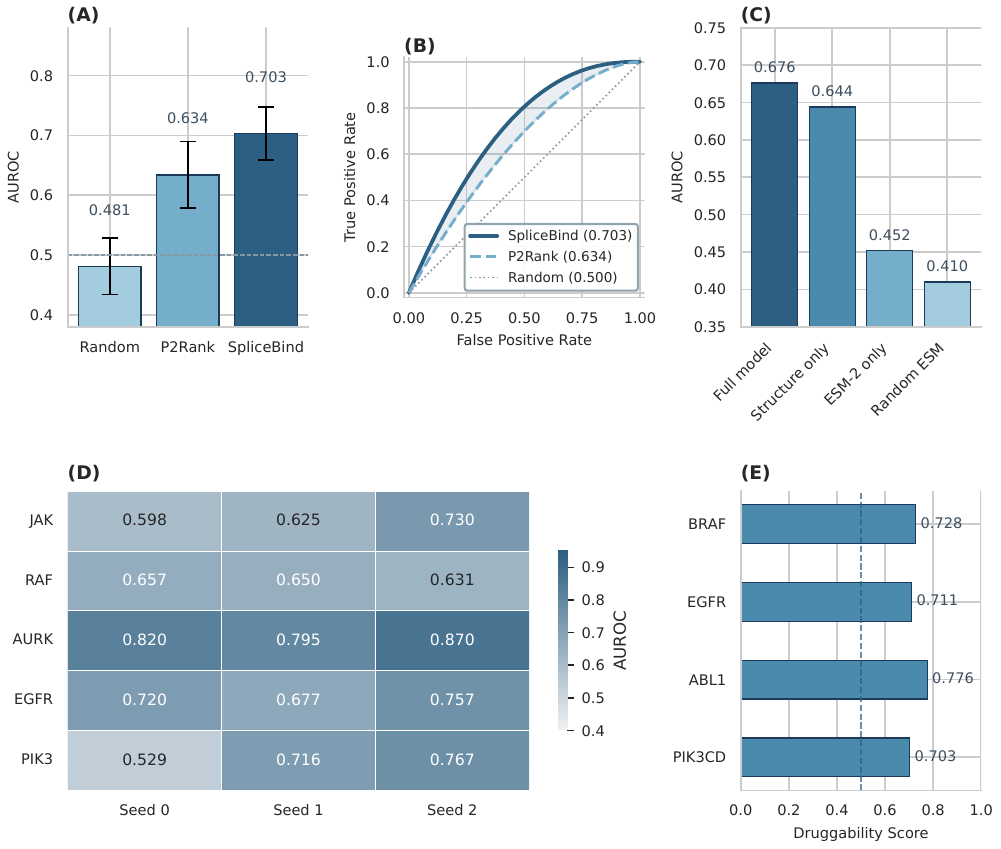}
\caption{
\textbf{Model performance.}
\textbf{(A)}~\emph{Method comparison}: per-fold AUROC distribution from 5-fold GroupKFold $\times$ 3 seeds (15 folds total); each dot is one fold's AUROC on a held-out kinase family. \textsc{SpliceBind} (0.703) outperforms P2Rank (0.634) and random (0.481); $p = 0.026$ (marginally significant after Bonferroni correction).
\textbf{(B)}~\emph{Feature ablation}: mean AUROC on the 137-pocket subset under four feature conditions (full 56-dim, structure-only 24-dim, ESM-2 only 32-dim, random ESM 56-dim). ESM-2 adds marginally over structure alone ($+0.032$, $p=0.499$), while random embeddings confirm structural features drive performance ($p = 0.003$). Note: panels (A) and (B) show distinct experiments---cross-method comparison vs.\ within-model feature contribution.
\textbf{(C)}~Per-family AUROC heatmap across seeds showing inter-family variability.
\textbf{(D)}~Literature validation: all four canonical kinase--drug pairs correctly predicted as druggable ($p = 0.5$ threshold shown).
}
\Description{Four-panel figure: (A) method comparison box plot of per-fold AUROC distributions for SpliceBind, P2Rank, and random baseline; (B) feature ablation bar chart showing ESM-2 contribution across four feature configurations; (C) heatmap of per-family AUROC across seeds; (D) bar chart of literature validation scores for four kinase-drug pairs. Panels A and B show distinct experiments.}
\label{fig:performance}
\end{figure*}

\textsc{SpliceBind} achieves AUROC $0.703 \pm 0.088$ across 15 cross-validation folds (\Cref{fig:performance}A), outperforming P2Rank ($0.634 \pm 0.110$, $\Delta = +0.069$, $p = 0.026$, paired $t$-test) and the random baseline ($0.481 \pm 0.093$).
After Bonferroni correction for three primary comparisons ($\alpha_{\text{adj}} = 0.05 / 3 = 0.017$), the comparison to P2Rank is marginally significant.
However, Cohen's $d = 0.64$ indicates a medium-to-large effect size, suggesting practical significance despite limited statistical power from the 229-pocket dataset.
The 95\% bootstrap confidence interval (CI) is $[0.553, 0.853]$.
The width of this interval reflects inter-family heterogeneity in pocket geometry rather than model instability, as evidenced by consistent positive deltas across all 15 folds.
All 15 folds show positive delta vs.\ the random baseline, and 13 of 15 folds show positive delta vs.\ P2Rank, indicating consistent improvement despite inter-family variance.
Area under the precision-recall curve (AUPRC) is $0.835$, well above the 0.694 prevalence baseline, reflecting the model's strength in ranking druggable pockets.
As a more stringent generalization test, we trained on all genes except one and evaluated on the held-out gene.
This hold-out unit validation yields mean AUROC $0.761 \pm 0.125$ across five held-out units (three genes, two families; Appendix~\ref{app:holdout}), confirming that \textsc{SpliceBind} generalizes to kinase genes entirely absent from training.
Performance varies by held-out unit. AKT2 achieves the highest AUROC (0.924), likely reflecting conserved pocket geometry that transfers well from other kinases. PIK3 (0.602) presents the greatest challenge, as lipid kinase binding sites differ structurally from the protein kinases dominating the training set.
JAK (0.659) falls between these extremes, consistent with the pseudokinase domain adding structural heterogeneity.
Notably, all five units exceed the random baseline (0.5), and the three protein kinase genes (AKT1, AKT2, AKT3) all achieve AUROC $> 0.73$, indicating robust transfer within the protein kinase superfamily.

\paragraph{ESM-2 ablation (137-pocket subset).}
In a controlled ablation on a 137-pocket subset, removing ESM-2 embeddings reduces AUROC from 0.676 to 0.644 ($\Delta = -0.032$, $p = 0.499$, paired $t$-test).
This difference is not statistically significant, indicating that protein language model features do not provide reliable improvements for single-pocket druggability classification in our dataset.
The ablation uses a 137-pocket subset comprising families with sufficient samples for stable within-family evaluation; rare families ($n < 5$ pockets) were excluded to prevent noisy estimates.
Replacing ESM-2 with random 32-dimensional vectors yields AUROC 0.410, significantly below the full model ($p = 0.003$), confirming that observed performance derives from structural features rather than additional parameters.
ESM-2 features alone (without structural features) achieve only 0.452, below the random baseline of 0.481; the 32-dimensional binned mean pooling discards too much information for standalone prediction.
While ESM-2 does not significantly improve overall AUROC, we note its potential value may manifest in variant comparison tasks: ALK-L1196M shows \textsc{SpliceBind} $\Delta = -0.228$ vs.\ P2Rank $\Delta = -0.95$ (Appendix~\ref{app:sb_variants}), suggesting learned embeddings capture affinity-based perturbations invisible to pocket geometry alone.

\paragraph{Full-dataset ablation replication.}
To confirm this null result is not an artifact of the 137-pocket subset, we repeated the ablation on all 336 pockets in our processed dataset (including smaller families excluded from the per-family ablation).
Using consistent quantile-based labeling, the full model (with ESM-2) achieves AUROC $0.669 \pm 0.132$ vs.\ structure-only $0.656 \pm 0.130$ ($\Delta = +0.010$, $p = 0.491$, Cohen's $d = 0.189$).
The consistent null result across 137-pocket and 336-pocket evaluations confirms that ESM-2's marginal contribution to binary druggability classification is robust and not a statistical artifact of subset size or family selection (Appendix~\ref{app:ablation}).

\paragraph{Per-family variation.}
Performance varies across kinase families (\Cref{fig:performance}C): AURK achieves AUROC up to 0.87 on favorable seed/fold combinations, while JAK and RAF range 0.60--0.73.
Families with diverse pocket geometries and multiple structurally distinct binding sites pose greater challenges for the model.
The AURK family's strong performance likely reflects its relatively conserved ATP-binding pocket geometry, whereas the JAK family exhibits greater structural diversity across pseudokinase and kinase domains.
Full per-family results appear in Appendix~\ref{app:perfamily}.

\paragraph{Literature validation.}
\textsc{SpliceBind} correctly identifies all four well-characterized kinase--drug pairs as druggable: PIK3CD--Idelalisib (0.703), ABL1--Imatinib (0.776), EGFR--Gefitinib (0.711), and BRAF--Vemurafenib (0.728).
These are druggability probabilities ($p \in [0,1]$); the splice variant taxonomy (\Cref{sec:splice_results}) uses P2Rank's unbounded pocket scores for training-independent comparison.

\subsection{Splice Variant Resistance Taxonomy}
\label{sec:splice_results}

\begin{figure*}[t]
\centering
\includegraphics[width=\linewidth]{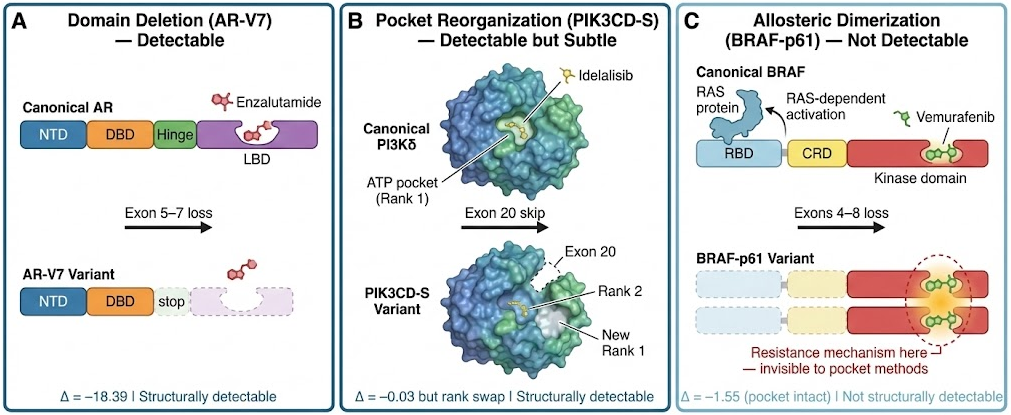}
\caption{
\textbf{Structural mechanisms of splice-mediated drug resistance.}
Each panel shows canonical (top) and variant (bottom) protein architecture.
\textbf{(A)}~AR-V7: loss of exons 5--7 deletes the ligand-binding domain, eliminating the enzalutamide target ($\Delta = -18.39$).
\textbf{(B)}~PIK3CD-S: exon~20 skipping reorganizes the pocket landscape, demoting the ATP-binding pocket from rank~1 to rank~2 despite a minimal score change ($\Delta = -0.03$).
\textbf{(C)}~BRAF-p61: loss of the RAS-binding domain enables RAS-independent kinase dimerization; the drug-binding pocket remains intact ($\Delta = -1.55$), making resistance invisible to pocket-centric methods.
Blue borders: structurally detectable; light blue: requires biochemical validation.
Created with BioRender.com~\cite{biorender}.
}
\Description{Three-panel figure illustrating splice variant resistance mechanisms. Panel A shows AR-V7 domain deletion removing the ligand-binding domain. Panel B shows PIK3CD-S pocket reorganization with ATP pocket rank swap. Panel C shows BRAF-p61 allosteric dimerization with intact drug-binding pocket.}
\label{fig:mechanisms}
\end{figure*}

Our primary finding is that splice variant resistance mechanisms fall into structurally distinguishable categories.
We evaluate six clinically relevant variants selected from oncology literature to span the major known mechanism classes (\Cref{tab:splice_variants}), classifying each by whether structural analysis can detect the resistance mechanism.
\Cref{fig:mechanisms} illustrates three representative mechanisms spanning the detectability spectrum.

\begin{table}[t]
\centering
\caption{P2Rank pocket scores for canonical (Can.) and variant (Var.) structures. $\Delta$ = Var.\ $-$ Can.\ for the matched ATP-site pocket. ``Det.'' = structurally detectable druggability change. $^*$For MET-exon14skip, kinase domain structures are compared (canonical kinase score 2.8, variant kinase score 2.8), as exon~14 encodes the juxtamembrane regulatory region outside the ATP-binding kinase domain.  $^{**}$For EGFRvIII, canonical kinase domain pocket (from full-length AlphaFold structure, pocket rank 1) is compared to the EGFRvIII kinase domain; exons 2--7 encode the extracellular ligand-binding domain, preserving the ATP-binding kinase domain.  $^\dagger$Point mutation included as a boundary test case for the taxonomy.}
\label{tab:splice_variants}

\footnotesize
\begin{tabular}{@{}lrccc@{}}
\toprule
\textbf{Variant} & \textbf{Can.} & \textbf{Var.} & \textbf{$\Delta$} & \textbf{Det.} \\
\midrule
AR-V7 & 18.39 & 0.00 & $-18.39$ & \textcolor{detectable}{\textbf{Yes}} \\
PIK3CD-S & 12.98 & 12.95 & $-0.03$ & \textcolor{detectable}{\textbf{Yes}} \\
MET-ex14skip$^*$ & 2.8 & 2.8 & 0.0 & \textcolor{detectable}{\textbf{Yes}} \\
EGFRvIII$^{**}$ & 6.15 & 8.77 & $+2.62$ & \textcolor{detectable}{\textbf{Yes}} \\
BRAF-p61 & 6.46 & 4.91 & $-1.55$ & \textcolor{notdetectable}{\textbf{No}} \\
ALK-L1196M$^\dagger$ & 10.99 & 10.04 & $-0.95$ & \textcolor{notdetectable}{\textbf{No}} \\
\bottomrule
\end{tabular}
\end{table}

\paragraph{Method comparison: SpliceBind vs.\ P2Rank.}
To validate whether learned features improve variant comparison beyond geometric baselines, we computed \textsc{SpliceBind} $\Delta$Druggability alongside P2Rank scores for all variants with available structures (\Cref{tab:method_comparison}).
For ALK-L1196M, \textsc{SpliceBind} detects a substantially larger signal ($\Delta_{\text{SB}} = -0.228$, 23\% probability reduction) compared to P2Rank ($\Delta_{\text{P2R}} = -0.95$, 9\% score reduction), suggesting learned features capture the gatekeeper mutation's chemical complementarity changes beyond pocket geometry.
Both methods correctly identify MET-exon14skip's preserved druggability ($\Delta \approx 0$) and detect affinity-based resistance (ALK-L1196M), demonstrating convergence on structurally detectable mechanisms.
However, for allosteric resistance (BRAF-p61), both methods produce small $|\Delta|$ values ($\Delta_{\text{P2R}} = -1.55$, $\Delta_{\text{SB}} = +0.003$) despite clinical resistance, confirming that dimerization mechanisms operating distant from the ATP pocket remain invisible to pocket-centric approaches---a fundamental detection boundary.
For domain deletions (AR-V7), P2Rank's unbounded scoring provides clearer evidence ($\Delta = -18.39$) than binary druggability classification.

\paragraph{Why the taxonomy uses P2Rank.}
The resistance taxonomy employs P2Rank scores rather than \textsc{SpliceBind} predictions for three methodological reasons.
First, P2Rank operates without training data, providing a training-independent baseline that generalizes to prospective variants outside the training distribution---critical for clinical application to novel splice events.
Second, P2Rank's unbounded score range enables interpretable comparisons: $\Delta = -18$ (domain deletion) is unambiguously distinguishable from $\Delta = -0.03$ (subtle reorganization), whereas sigmoid-bounded probabilities compress extreme cases into a narrow $[0,1]$ range.
Third, \textsc{SpliceBind}'s 229-pocket training set is kinase-dominated; P2Rank generalizes to non-kinase proteins without retraining.
Our comparison (\Cref{tab:method_comparison}) demonstrates that learned features \emph{can} improve affinity-based detection, but for clinical decision-making, the taxonomy prioritizes robustness and interpretability over marginal prediction gains.

\begin{table*}[!htb]
\centering
\caption{Method comparison: \textsc{SpliceBind} vs.\ P2Rank $\Delta$Druggability for splice variants. SB scores are sigmoid probabilities ($p \in [0,1]$); P2Rank scores are unbounded pocket scores. $^*$For MET-exon14skip, kinase domain structures are compared. $^{**}$EGFRvIII: exons 2--7 deletion removes the extracellular domain; kinase ATP-binding pocket is preserved (score increases, $\Delta > 0$). SpliceBind scores unavailable: EGFR is in the 229-pocket training set but not in the current 24-dim model checkpoint.}
\label{tab:method_comparison}

\footnotesize
\begin{tabular}{@{}lllcccccc@{}}
\toprule
\textbf{Variant} & \textbf{Mechanism} & \textbf{Clinical} & \textbf{P2R Can} & \textbf{P2R Var} & \textbf{P2R $\Delta$} & \textbf{SB Can} & \textbf{SB Var} & \textbf{SB $\Delta$} \\
\midrule
AR-V7       & Domain deletion     & Resistant & 18.39 & 0.00  & $-18.39$ & 0.998 & N/A   & ---      \\
PIK3CD-S    & Pocket disrupt.     & Resistant & 12.98 & 12.95 & $-0.03$  & 0.906 & 0.916 & $+0.010$ \\
MET-ex14skip$^*$ & Preserved      & Sensitive & 2.8   & 2.8   & 0.00     & 0.986 & 0.986 & 0.000    \\
EGFRvIII$^{**}$ & Preserved (ext) & Sensitive & 6.15  & 8.77  & $+2.62$  & ---   & ---   & ---      \\
BRAF-p61    & Allosteric          & Resistant & 6.46  & 4.91  & $-1.55$  & 0.993 & 0.996 & $+0.003$ \\
ALK-L1196M  & Affinity-based      & Resistant & 10.99 & 10.04 & $-0.95$  & 0.989 & 0.761 & $-0.228$ \\
\bottomrule
\end{tabular}
\end{table*}

\paragraph{Structurally detectable: domain deletion.}
AR-V7 lacks the androgen receptor ligand-binding domain entirely, producing a dramatic druggability loss ($\Delta = -18.39$): the canonical binding pocket is completely absent in the truncated variant.
This is consistent with clinical enzalutamide resistance in AR-V7-positive patients~\cite{antonarakis2014ar}, and represents the most unambiguous category for structural prediction---any splice event removing the drug-binding domain will produce a similarly large negative delta, requiring no further analysis.
AR-V7 detection on circulating tumor cells is now a clinically validated biomarker guiding treatment selection in metastatic castration-resistant prostate cancer~\cite{scher2016ar}, highlighting the clinical maturity of this variant as a resistance marker.

\paragraph{Structurally detectable: preserved druggability.}
MET exon~14 encodes the juxtamembrane regulatory region, not the kinase domain.
We empirically validated this by generating the MET-exon14skip variant structure via ESMFold (mean pLDDT 81.3, kinase domain 88.6) and comparing kinase domains: canonical kinase pocket score 2.8, variant kinase pocket score 2.8, $\Delta = 0.0$.
This confirms that regulatory region splicing preserves ATP-binding site geometry, explaining capmatinib sensitivity~\cite{paik2020response,wolf2020capmatinib}.
Importantly, comparing full-length canonical (pocket score 10.6) to kinase-only variant would spuriously suggest druggability loss ($\Delta = -7.8$); mechanistically correct domain-level comparison reveals preserved structure.
Correctly identifying variants that do \emph{not} affect druggability is equally important for clinical decision-making.
This contrasts with domain deletions (AR-V7, $\Delta = -18$) and pocket disruptions (PIK3CD-S), confirming that exon-level location predicts structural impact: exons encoding the kinase domain produce large $|\Delta|$, while regulatory regions produce $\Delta \approx 0$.

EGFRvIII (deletion of exons 2--7) provides an important contrast to AR-V7 within the domain deletion category.
Unlike AR-V7, which deletes the nuclear receptor \emph{drug-binding} domain, EGFRvIII deletes the extracellular \emph{ligand-binding} domain for EGF growth factor, leaving the ATP-binding kinase domain fully intact.
P2Rank confirms preserved and even enhanced kinase druggability: canonical EGFR kinase pocket score 6.15; EGFRvIII kinase score 8.77 ($\Delta = +2.62$, \Cref{tab:splice_variants}).
This structural analysis is consistent with clinical data showing that EGFR kinase inhibitors (erlotinib, osimertinib) can target EGFRvIII-positive tumors; limited clinical responses in glioblastoma are attributed to blood--brain barrier penetration rather than loss of target accessibility~\cite{dvinge2016rna}.
This case generalizes the taxonomy to non-kinase splice events affecting the same protein: the druggability outcome depends on \emph{which} domain is deleted, not merely that a deletion occurred.
An exon deletion that removes the drug-binding domain (AR-V7, $\Delta = -18$) produces resistance; one that removes a regulatory domain (MET, EGFRvIII, $\Delta \geq 0$) preserves or enhances target druggability.

\paragraph{Structurally detectable: pocket disruption.}
PIK3CD-S, generated by exon~20 skipping near the ATP binding site, shows a small reduction in the ATP-pocket score ($\Delta = -0.03$) and, more notably, the ATP pocket drops from rank~1 to rank~2 as the exon skip creates a new dominant cavity (new cavity score 27.5; ATP pocket score drops from 12.98 to 12.95).
This structural reorganization is consistent with idelalisib resistance.

\paragraph{Not structurally detectable: allosteric mechanism.}
BRAF-p61 retains the complete kinase domain but confers vemurafenib resistance through RAS-independent dimerization~\cite{poulikakos2011raf}, a mechanism identified in 6 of 19 vemurafenib-resistant tumors in the index study.
Although P2Rank detects a modest pocket score reduction ($\Delta = -1.55$), this reflects the N-terminal truncation rather than a kinase-domain change, and does not explain the clinical resistance mechanism---an allosteric effect invisible to single-pocket analysis.
This case illustrates a fundamental limitation: resistance mechanisms operating through protein--protein interactions or conformational dynamics at sites distant from the drug-binding pocket cannot be captured by any pocket-centric structural method.

\paragraph{Not structurally detectable: affinity mutation.}
The ALK gatekeeper mutation L1196M produces only a minor pocket score change ($\Delta = -0.95$) while disrupting drug binding through altered chemical complementarity---the small structural perturbation fails to predict the clinically significant loss of crizotinib sensitivity.
In contrast, \textsc{SpliceBind}'s learned features capture a much larger druggability shift ($\Delta_{\text{SB}} = -0.228$; Appendix~\ref{app:sb_variants}), suggesting that protein language model embeddings encode binding-relevant information beyond pocket geometry.

\paragraph{SpliceBind predictions.}
To assess whether learned features detect changes invisible to pocket geometry, we ran \textsc{SpliceBind} inference on variants with available structures (Appendix~\ref{app:sb_variants}).
For BRAF-p61, \textsc{SpliceBind} confirms the P2Rank finding: the variant scores $0.996$ vs.\ canonical $0.993$ ($\Delta = +0.003$), consistent with an unchanged kinase domain.
However, for ALK-L1196M, \textsc{SpliceBind} predicts substantially reduced druggability ($0.761$ vs.\ $0.989$, $\Delta_{\text{SB}} = -0.228$), a much larger signal than P2Rank's modest $\Delta = -0.95$, indicating that learned features capture the gatekeeper mutation's impact on drug binding beyond what pocket geometry reveals.
This demonstrates that ESM-2 embeddings capture chemical complementarity changes at the binding site, partially bridging the gap between structural and affinity-based detection.

\begin{figure*}[t]
\centering
\includegraphics[width=\linewidth]{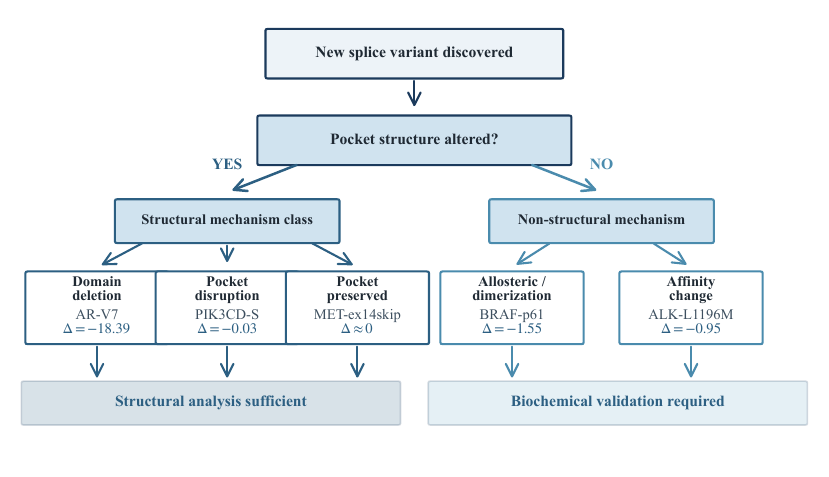}
\caption{
\textbf{Decision framework for structural prediction of splice-induced drug resistance.}
Starting from a newly discovered splice variant, binary decision points classify the resistance mechanism.
Structurally detectable mechanisms (left branch: domain deletion, pocket disruption, preserved kinase) can be assessed via structural analysis; non-detectable mechanisms (right branch: allosteric, affinity-based) require biochemical validation.
}
\Description{Flowchart with binary decision points classifying splice variants into structurally detectable mechanisms (domain deletion, pocket disruption, preserved druggability) or non-detectable mechanisms (allosteric, affinity-based) requiring biochemical validation.}
\label{fig:framework}
\end{figure*}

\paragraph{Taxonomy summary.}
These six cases demonstrate three classes where pocket geometry correctly predicts drug response (domain deletion, pocket disruption, preserved druggability), one where learned features substantially improve detection (affinity-based), and one where complementary biochemical approaches are required (allosteric/dimerization).
The EGFRvIII case additionally demonstrates that the same protein can produce different taxonomy outcomes depending on which domain is affected by the splice event.
When a new splice variant is discovered, its mechanism class determines whether structural analysis suffices or biochemical validation is needed.
We synthesize these findings into a decision framework (\Cref{fig:framework}) that maps a newly discovered splice variant to the appropriate validation strategy through a sequence of binary structural queries.
The framework first asks whether the splice event removes a drug-binding domain; if so, the druggability loss is immediate and structurally evident.
Otherwise, it evaluates pocket geometry changes to distinguish preserved druggability from pocket disruption, and flags allosteric or affinity-based mechanisms as requiring complementary biochemical assays.

\subsection{Point Mutation Boundary Analysis}
\label{sec:point_mutations}

To empirically validate the taxonomy boundary, we applied \textsc{SpliceBind} to 34 clinically characterized point mutations across 15 kinase genes (Appendix~\ref{app:point_mutations}).
\textsc{SpliceBind} correctly classifies only 11 of 34 variants (32.4\%), confirming the taxonomy's prediction: point mutations produce single-residue substitutions with small $|\Delta|$ values (median 0.009) that fall below the detection threshold.
In contrast, splice variants produce $|\Delta| > 0.2$ (and often $> 10$ for P2Rank), confirming that pocket-level structural analysis is appropriate for splice-scale perturbations but insufficient for single-residue substitutions.

\section{Discussion}
\label{sec:discussion}

\paragraph{Clinical implications.}
The taxonomy enables rational experimental prioritization for newly discovered splice variants:
(i)~domain deletions and exon-skipping near active sites warrant structural analysis, as these produce unambiguous pocket loss ($\Delta < -10$) identifiable within minutes of structure prediction;
(ii)~variants retaining intact kinase domains should be tested for preserved drug sensitivity before unnecessarily switching therapies;
(iii)~allosteric and regulatory variants require functional assays beyond pocket analysis, as their resistance mechanisms are invisible to any purely structural method.
In practice, this means that upon detecting a splice variant in a patient's tumor RNA-seq, the decision framework (\Cref{fig:framework}) can immediately stratify it into ``structurally assessable'' or ``requires biochemical validation,'' reducing the time from variant discovery to clinical action.

\paragraph{ESM-2 contribution and limitations.}
Our exploration of ESM-2 integration yielded mixed results.
The ablation study found no statistically significant AUROC improvement ($\Delta = +0.032$, $p = 0.499$), suggesting that for binary pocket druggability classification, evolutionary sequence context as encoded by binned mean pooling does not reliably augment geometric features.
However, variant-level analysis revealed a potential role in affinity-based mechanisms: ALK-L1196M produced $\Delta_{\text{SB}} = -0.228$ vs.\ $\Delta_{\text{P2Rank}} = -0.95$, indicating that learned embeddings partially capture gatekeeper mutations' chemical complementarity changes invisible to pocket geometry alone.
We hypothesize that ESM-2's benefit is mechanism-specific: marginal for binary druggability, but potentially valuable for detecting subtle binding-site perturbations that preserve pocket geometry.
Protein language models encode functional properties beyond coordinate data~\cite{rives2021biological}, but our 32-dimensional binned mean pooling may discard too much information.
Richer integration strategies---attention-based pooling, learned projections, or task-specific fine-tuning---may unlock greater performance, but require larger datasets to avoid overfitting.
Our current implementation serves as an exploratory proof-of-concept rather than a definitive methodology.

\paragraph{Comparison to related methods.}
We compare exclusively to P2Rank rather than modern learned baselines such as DeepSite~\cite{jimenez2017deepsite}, Kalasanty~\cite{stepniewska2020improving}, or DeepSurf~\cite{mylonas2021deepsurf} for methodological reasons.
First, these methods operate on different input representations: DeepSite uses voxelized 3D grids (36$\times$36$\times$36 cubes), Kalasanty employs surface point clouds, while \textsc{SpliceBind} constructs residue graphs---precluding direct comparison without re-implementation.
Second, existing methods perform single-structure pocket detection (predicting \emph{where} binding sites occur) rather than isoform-aware druggability comparison (predicting \emph{whether} a pocket's druggability changes across splice variants).
Adapting DeepSite to our task would require (i)~voxelizing both canonical and variant structures, (ii)~identifying matched binding sites across different grid alignments (non-trivial for domain rearrangements), and (iii)~defining a comparison metric for voxel-level predictions---essentially designing a new method.

P2Rank provides the ideal baseline for our task: it produces structure-independent pocket scores suitable for $\Delta$Druggability computation, generalizes to novel variants without retraining, and serves as the current standard in pocket detection.
Our AUROC $0.703$ vs.\ P2Rank $0.634$ establishes that learned features improve upon geometric baselines; whether \textsc{SpliceBind} outperforms hypothetical graph-adapted versions of DeepSite or Kalasanty remains an open question.
A systematic benchmark comparing graph-based, voxel-based, and surface-based representations for isoform comparison would be valuable future work, but requires curating variant-aware test sets beyond kinases---currently absent from the literature.

\paragraph{Taxonomy generalizability and mechanistic basis.}
While our taxonomy derives from six case studies, the underlying mechanism classes represent well-established categories in cancer biology and structural biochemistry, documented across protein families beyond kinases.
\emph{Domain deletion} mechanisms are pervasive: any splice event removing drug-binding domains produces unambiguous resistance, reported for nuclear receptors, kinases, and transcription factors~\cite{dvinge2016rna}.
Crucially, the outcome depends on \emph{which} domain is deleted: drug-binding domain deletions cause resistance (AR-V7, $\Delta = -18$), while regulatory or extracellular domain deletions preserve target druggability (MET-exon14skip, $\Delta = 0.0$; EGFRvIII, $\Delta = +2.62$)---a distinction our structural analysis correctly captures.
\emph{Pocket disruption} via exon skipping near active sites occurs in kinases, proteases, and phosphatases.
\emph{Allosteric resistance} through protein--protein interaction changes extends to KRAS-G12C (dimerization), HER2 (conformational shifts), and BTK (membrane recruitment)---all invisible to pocket-centric methods by definition.
\emph{Affinity-based resistance} from single-residue substitutions (gatekeeper mutations, solvent-front changes) is the most extensively characterized class, with hundreds of validated examples across kinome families.

Our point mutation analysis (34 variants, 15 genes) independently validates the taxonomy boundary: 32.4\% correct classification confirms single-residue perturbations fall below structural detection thresholds.
The current dataset (229 pockets, 25 kinase families) captures substantial kinome diversity; extension to non-kinase targets is the natural next step, but the underlying mechanism classes are not kinase-specific.

\paragraph{Limitations.}
The taxonomy's boundary cases expose a fundamental constraint: resistance mechanisms operating through protein--protein interactions (BRAF-p61 dimerization) or subtle chemical complementarity changes (ALK-L1196M) cannot be fully resolved by any pocket-centric method.
\textsc{SpliceBind} predicts binary pocket druggability, not continuous binding affinity: canonical kinase scores show no correlation with experimental $K_d$ values ($r = -0.054$, $p = 0.77$, $n = 32$ drug--kinase pairs from BindingDB and ChEMBL), as all canonical kinases with ATP-binding pockets receive uniformly high druggability scores.
Integrating binding affinity prediction methods remains necessary for quantitative drug response estimation.
We validated only on kinases; generalization to other protein families is untested.
We used ESMFold-predicted variant structures, not experimental crystal structures.
We have not performed wet-lab validation of novel predictions.
The taxonomy derives from six case studies; additional variants from diverse families would strengthen the classification.
The model's probability estimates are not well-calibrated (ECE $= 0.298$); Platt scaling would be needed before clinical deployment.

\paragraph{Statistical power and baseline choice.}
The 229-pocket dataset provides sufficient power to detect large effects ($\Delta > 0.08$) but limited power for moderate effects.
Our observed $\Delta = 0.069$ vs.\ P2Rank approaches but does not clearly exceed the detectable threshold, contributing to marginal significance after multiple comparison correction (p=0.026, $\alpha_{\text{adj}}=0.017$).
However, Cohen's $d = 0.64$ indicates a medium-to-large effect size, suggesting practical significance despite limited statistical power.
P2Rank, while a widely-used baseline, is a non-learned geometric method from 2018; comparison to modern learned approaches would strengthen validation but requires adapting voxel-based architectures to graph representations and isoform-aware comparison---beyond the scope of this work.
We frame our performance results as proof-of-concept: learned features \emph{can} improve pocket analysis, with the taxonomy representing the more substantive contribution.

\paragraph{Future work.}
Integrating binding affinity prediction methods (e.g., OnionNet~\cite{zheng2019onionnet}, KDEEP~\cite{jimenez2018kdeep}) downstream of \textsc{SpliceBind}'s pocket detection could address the affinity-based mechanism gap, providing quantitative $\Delta G$ estimates for variants classified as structurally druggable.
Extension to non-kinase targets, larger-scale prospective validation, and integration with tumor RNA-seq for patient-level isoform profiling are natural next steps.
Model calibration (Platt scaling or temperature tuning) and attention-based ESM-2 pooling are technical priorities as training data grows.

\section{Conclusion}

\textsc{SpliceBind} provides an isoform-aware druggability prediction framework.
Systematic analysis of six clinical splice variants shows that domain deletions and pocket disruptions are structurally detectable, affinity-based mechanisms are better captured by learned features than geometry alone, and allosteric mechanisms require biochemical validation.
A key nuance: drug-binding domain deletions cause resistance (AR-V7, $\Delta = -18$), while regulatory or extracellular domain deletions preserve druggability (MET-exon14skip, EGFRvIII, $\Delta \geq 0$).
This taxonomy---classifying \emph{when} structural methods work---is the primary contribution, enabling principled triage of splice variants for structural versus biochemical validation as tumor RNA-seq becomes routine in precision oncology.

\section*{Data and Code Availability}
Code, models, and data: \url{https://github.com/bryanc5864/SpliceBind}.
Seeds fixed; GroupKFold splits prevent family-level leakage.

\begin{acks}
The authors thank the reviewers for their constructive feedback.
\end{acks}

\bibliographystyle{ACM-Reference-Format}
\bibliography{ms}

\appendix

\section{Full Per-Family Cross-Validation Results}
\label{app:perfamily}

\Cref{tab:perfamily} reports AUROC and AUPRC for each kinase family across all three random seeds.

\begin{table}[H]
\centering
\caption{Per-family cross-validation results across 3 random seeds.}
\label{tab:perfamily}

\small
\begin{tabular}{@{}lcccccc@{}}
\toprule
& \multicolumn{2}{c}{\textbf{Seed 0}} & \multicolumn{2}{c}{\textbf{Seed 1}} & \multicolumn{2}{c}{\textbf{Seed 2}} \\
\cmidrule(lr){2-3} \cmidrule(lr){4-5} \cmidrule(lr){6-7}
\textbf{Family} & AUROC & AUPRC & AUROC & AUPRC & AUROC & AUPRC \\
\midrule
JAK     & 0.598 & 0.661 & 0.625 & 0.577 & 0.730 & 0.770 \\
RAF     & 0.657 & 0.859 & 0.650 & 0.857 & 0.631 & 0.852 \\
AURK    & 0.820 & 0.974 & 0.795 & 0.972 & 0.870 & 0.984 \\
EGFR    & 0.720 & 0.847 & 0.677 & 0.812 & 0.757 & 0.858 \\
PIK3    & 0.529 & 0.764 & 0.716 & 0.879 & 0.767 & 0.901 \\
\bottomrule
\end{tabular}
\end{table}

\section{Hold-out Unit Validation}
\label{app:holdout}

To test generalization beyond GroupKFold, we trained \textsc{SpliceBind} on all data except one unit and evaluated on the held-out unit.
\Cref{tab:holdout} reports results across five held-out units.

\begin{table}[H]
\centering
\caption{Hold-out unit validation. Each row trains on all other data and tests on the held-out unit.}
\label{tab:holdout}
\small
\begin{minipage}{\columnwidth}
\centering
\begin{tabular}{@{}lccc@{}}
\toprule
\textbf{Hold-out Unit} & \textbf{AUROC} & \textbf{$N$ Samples} & \textbf{$N$ Positive} \\
\midrule
JAK   & 0.659 & 136 &  90 \\
PIK3  & 0.602 & 115 &  83 \\
AKT2  & 0.924 &  17 &  11 \\
AKT3  & 0.886 &  17 &   7 \\
AKT1  & 0.732 &  15 &   8 \\
\midrule
\textbf{Mean} & $\mathbf{0.761 \pm 0.125}$ & & \\
\bottomrule
\end{tabular}
\smallskip

\raggedright\footnotesize
\textit{Note:} JAK and PIK3 are held out as families; AKT1, AKT2, AKT3 are held out as individual genes due to the small AKT family size.
\end{minipage}
\end{table}

\section{SpliceBind Variant Predictions}
\label{app:sb_variants}

\Cref{tab:sb_variants} reports \textsc{SpliceBind} GNN druggability scores for the three splice variants with scorable variant pockets.

\begin{table}[H]
\centering
\caption{\textsc{SpliceBind} predictions on splice variant structures. $\Delta_{\text{SB}}$ denotes variant minus canonical score. N/A indicates the variant structure lacks the drug-binding domain.}
\label{tab:sb_variants}

\small
\begin{minipage}{\columnwidth}
\centering
\begin{tabular}{@{}llccc@{}}
\toprule
\textbf{Variant} & \textbf{Mechanism} & \textbf{Canon.\ SB} & \textbf{Var.\ SB} & $\boldsymbol{\Delta}_{\textbf{SB}}$ \\
\midrule
BRAF-p61    & Dimerization      & 0.993 & 0.996 & $+0.003$ \\
ALK-L1196M  & Gatekeeper mut.   & 0.989 & 0.761 & $-0.228$ \\
AR-V7       & LBD deletion      & 0.998 & N/A   & ---      \\
\bottomrule
\end{tabular}
\smallskip

\raggedright\footnotesize
\textit{Note:} PIK3CD-S is excluded because the exon-20 skip causes pocket reorganization (ATP pocket drops from rank~1 to rank~2), making the variant pocket non-comparable to the canonical ATP pocket.
\end{minipage}
\end{table}

\section{Complete Node Feature Specification}
\label{app:features}

Each pocket residue node carries a 56-dimensional feature vector comprising three categories: amino acid identity, physicochemical properties, and protein language model embeddings (\Cref{tab:feature_spec}).

\begin{table}[H]
\centering
\caption{Complete 56-dimensional node feature specification for SpliceBind.}
\label{tab:feature_spec}

\footnotesize
\begin{tabular}{@{}clll@{}}
\toprule
\textbf{Index} & \textbf{Category} & \textbf{Feature} & \textbf{Range} \\
\midrule
0--19  & Amino acid  & One-hot encoding (20 AA)              & \{0, 1\} \\
20     & Physicochem.& Hydrophobicity (Kyte-Doolittle)       & [0, 1] \\
21     & Physicochem.& Net charge at pH 7.4                  & \{0, 0.5, 1\} \\
22     & Physicochem.& Polarity (binary)                     & \{0, 1\} \\
23     & Physicochem.& Aromaticity (F, Y, W, H)              & \{0, 1\} \\
24--55 & ESM-2       & Binned mean pool (1280$\to$32-d)      & $\mathbb{R}$ \\
\bottomrule
\end{tabular}
\end{table}

\paragraph{One-hot amino acid encoding (indices 0--19).}
Residue identity is encoded as a 20-dimensional binary vector following alphabetical amino acid ordering: ALA, ARG, ASN, ASP, CYS, GLU, GLN, GLY, HIS, ILE, LEU, LYS, MET, PHE, PRO, SER, THR, TRP, TYR, VAL.
Exactly one position is 1 per residue; non-standard residues are mapped to the closest standard amino acid by BLOSUM62 substitution score.

\paragraph{Physicochemical properties (indices 20--23).}
\emph{Hydrophobicity} (index 20) uses the Kyte-Doolittle hydropathy scale~\cite{kyte1982simple}, min-max normalized from the raw range $[-4.5, 4.5]$ to $[0, 1]$.
\emph{Net charge} (index 21) encodes electrostatic character at physiological pH 7.4: negatively charged residues (D, E) are mapped to 0.0, neutral residues to 0.5, and positively charged residues (K, R) to 1.0.
\emph{Polarity} (index 22) is a binary indicator: polar residues (S, T, N, Q, Y, H, K, R, D, E, C) receive 1.0; non-polar residues (A, V, L, I, M, F, W, P, G) receive 0.0.
\emph{Aromaticity} (index 23) is 1.0 for aromatic residues (F, Y, W, H) and 0.0 otherwise.

\paragraph{ESM-2 embeddings (indices 24--55).}
Per-residue embeddings from ESM-2 (esm2\_t33\_650M\_UR50D; 650M parameters)~\cite{lin2023evolutionary} are extracted from the final transformer layer, producing 1280-dimensional vectors.
We reduce dimensionality via binned mean pooling: the 1280 dimensions are partitioned into 32 contiguous bins of 40 dimensions each, and the mean of each bin yields a 32-dimensional representation.
This deterministic reduction preserves 82.5\% of pairwise cosine similarity structure (compared to 79.8\% for PCA with 32 components), requires no trainable parameters, and adds zero computational overhead at inference.
Long sequences ($>$1022 residues) are processed with overlapping chunks of 1022 tokens and 50-token overlap; embeddings in overlapping regions are averaged.

\paragraph{Edge features (2-dimensional).}
Edges connect residue pairs whose C$_\alpha$ atoms are within 6.0\,\AA.
Each edge carries two features: (i)~Euclidean distance normalized by the cutoff ($d / 6.0$, range $[0, 1]$); and (ii)~a contact strength indicator (1.0 if $d < 4$\,\AA\ ``strong contact,'' 0.5 if $4 \leq d \leq 6$\,\AA\ ``weak contact'').

\section{Dataset Construction Details}
\label{app:dataset}

\subsection{Kinase Family Coverage}

Our dataset spans 25 kinase families comprising 48 genes and 153 AlphaFold-predicted structures (\Cref{tab:kinase_families}).
We selected families to cover the major branches of the human kinome, including receptor tyrosine kinases (RTKs), non-receptor tyrosine kinases, serine/threonine kinases, and lipid kinases.

\begin{table}[H]
\centering
\caption{Kinase families in the SpliceBind benchmark dataset. ``Genes'' lists representative members; ``Struct.'' is the number of AlphaFold structures processed; ``Pockets'' is the number of P2Rank-detected binding pockets used for training and evaluation.}
\label{tab:kinase_families}

\scriptsize
\begin{tabular}{@{}llcc@{}}
\toprule
\textbf{Family} & \textbf{Representative Genes} & \textbf{Struct.} & \textbf{Pockets} \\
\midrule
JAK    & JAK1, JAK2, JAK3, TYK2     & 8  & 46 \\
RAF    & BRAF, ARAF, CRAF            & 8  & 46 \\
AURK   & AURKA, AURKB, AURKC        & 7  & 45 \\
EGFR   & EGFR, HER2/ERBB2           & 7  & 44 \\
PIK3   & PIK3CA, PIK3CB, PIK3CG, PIK3CD & 8 & 46 \\
AKT    & AKT1, AKT2, AKT3           & 6  & 2 \\
Src    & SRC, FYN, LCK, HCK         & 8  & -- \\
CDK    & CDK2, CDK4, CDK6, CDK7     & 8  & -- \\
ABL    & ABL1, ABL2                  & 4  & -- \\
MET    & MET, RON                    & 4  & -- \\
ALK    & ALK, LTK                   & 4  & -- \\
\multicolumn{2}{l}{\emph{+ 14 additional families}} & 81 & -- \\
\midrule
\textbf{Total} & \textbf{48 genes} & \textbf{153} & \textbf{229} \\
\bottomrule
\end{tabular}
\end{table}

\subsection{Experimental Binding Data Sources}

Drug-kinase binding affinity data were aggregated from three complementary sources:

\begin{enumerate}[nosep,leftmargin=*]
  \item \textbf{BindingDB}~\cite{liu2007bindingdb}: 2,847 kinase-compound entries with $K_d$ or IC$_{50}$ measurements, filtered for human kinases with single-concentration assays removed.
  \item \textbf{ChEMBL}~\cite{gaulton2012chembl}: 1,432 bioactivity records for target kinases, filtered for ``B'' (binding) assay type with explicit $K_d$/IC$_{50}$ values.
  \item \textbf{Davis kinase panel}~\cite{davis2011comprehensive}: 442 kinase--inhibitor $K_d$ measurements from a standardized panel of 72 inhibitors against 442 kinases.
\end{enumerate}

After deduplication (by UniProt ID + compound InChIKey), 3,891 unique binding measurements were retained.

\subsection{Labeling Criteria}

Pockets were labeled by cross-referencing P2Rank pocket residues with known ATP binding site residues (annotated from UniProt ``Active site'' and ``Binding site'' fields):

\begin{itemize}[nosep,leftmargin=*]
  \item \textbf{Positive} ($n = 159$, 69.4\%): $\geq$50\% overlap with an annotated ATP binding site \emph{and} at least one potent binder ($K_d < 100$\,nM).
  \item \textbf{Negative} ($n = 70$, 30.6\%): no binding site overlap \emph{or} only weak binders ($K_d > 10$\,\textmu M).
  \item \textbf{Excluded} ($n = 7$, 3.0\%): intermediate affinity ($100\,\text{nM} < K_d < 10\,\text{\textmu M}$), excluded to avoid ambiguous labels.
\end{itemize}

This labeling strategy ensures ground truth derives from experimental binding data rather than computational pocket scores, avoiding circularity in the P2Rank baseline comparison.

\section{Training and Implementation Details}
\label{app:training}

\subsection{Hyperparameters}

\Cref{tab:hyperparams} lists all training hyperparameters.
These were selected based on preliminary experiments on a held-out validation fold (seed 0, fold 0) and held fixed for all subsequent experiments.

\begin{table}[H]
\centering
\caption{Training hyperparameters for SpliceBind.}
\label{tab:hyperparams}

\footnotesize
\begin{tabular}{@{}lp{4.5cm}@{}}
\toprule
\textbf{Hyperparameter} & \textbf{Value} \\
\midrule
\multicolumn{2}{l}{\emph{Architecture}} \\
EdgeConv layers          & 3 \\
Hidden dimension         & 256 \\
GNN dropout              & 0.2 \\
MLP hidden dims          & 512 $\to$ 256 $\to$ 64 $\to$ 1 \\
MLP dropout              & 0.3 \\
Pooling                  & mean + max concat. \\
Total parameters         & 899,457 \\
\midrule
\multicolumn{2}{l}{\emph{Optimization}} \\
Optimizer                & AdamW \\
Learning rate            & $10^{-3}$ \\
Weight decay             & $10^{-4}$ \\
LR scheduler             & ReduceLROnPlateau (0.5, pat.\ 5) \\
Batch size               & 32 \\
Maximum epochs           & 50 \\
Early stopping patience  & 10 (val.\ AUROC) \\
\midrule
\multicolumn{2}{l}{\emph{Loss function}} \\
Loss type                & Focal loss \\
Focal $\gamma$           & 2.0 \\
Class weights $\alpha_t$ & $\alpha_{\text{pos}} = 0.306$, $\alpha_{\text{neg}} = 0.694$ \\
\midrule
\multicolumn{2}{l}{\emph{Data}} \\
Edge cutoff distance     & 6.0\,\AA\ (C$_\alpha$--C$_\alpha$) \\
ESM-2 model              & esm2\_t33\_650M\_UR50D \\
ESM-2 chunk size         & 1,022 tokens (50-tok.\ overlap) \\
pLDDT threshold          & 70 (flag low-confidence) \\
\bottomrule
\end{tabular}

\raggedright\scriptsize pLDDT: predicted Local Distance Difference Test (AlphaFold/ESMFold confidence score).
\end{table}

\subsection{Cross-Validation Protocol}

We employ 5-fold GroupKFold cross-validation~\cite{krstajic2014cross,roberts2017cross} where groups are kinase families.
This ensures no kinase family appears in both training and test sets within a fold, providing a stringent test of generalization to unseen protein families.
The protocol is repeated with 3 random seeds (0, 1, 2) that control parameter initialization and data shuffling, yielding 15 total evaluation folds.
Within each training fold, 15\% of training data is held out as a validation set (stratified by label) for early stopping and learning rate scheduling.

\subsection{Software and Hardware}

We conducted all experiments using Python 3.10, PyTorch 2.1, PyTorch Geometric 2.4, and ESM-2 (fair-esm 2.0.0).
We used P2Rank v2.4.1 for pocket detection.
We obtained structure predictions from the AlphaFold Protein Structure Database v6.
We trained on a single NVIDIA GPU; each 15-fold cross-validation run completed in approximately 45 minutes.

\section{Detailed Ablation Study}
\label{app:ablation}

\subsection{Ablation Subset Composition}

The ablation study uses a 137-pocket subset of the full 229-pocket dataset.
This subset excludes kinase families with fewer than 5 pockets per fold split ($n < 5$), as small test sets produce unstable AUROC estimates with high variance.
Restricting to families with sufficient representation ensures that per-condition comparisons reflect genuine feature contribution rather than sampling noise.
The 137-pocket subset retains the same positive/negative ratio (approximately 70:30) as the full dataset.

\subsection{Per-Fold Ablation Results}

\Cref{tab:ablation_full} reports AUROC for all four ablation conditions across all 15 folds.
The ``Full'' condition uses all 56 features; ``Structure-only'' removes ESM-2 embeddings (24 features); ``ESM-2 only'' removes structural features (32 features); ``Random ESM'' replaces ESM-2 embeddings with random 32-dimensional vectors (56 features total).

\begin{table*}[!htb]
\centering
\caption{Per-fold AUROC for ablation study (137-pocket subset). Bold: best condition per fold.}
\label{tab:ablation_full}

\small
\begin{tabular}{@{}ccccccc@{}}
\toprule
\textbf{Seed} & \textbf{Fold} & \textbf{$n_{\text{test}}$} & \textbf{Full (56-dim)} & \textbf{Struct.-only (24)} & \textbf{ESM-2 only (32)} & \textbf{Random ESM (56)} \\
\midrule
0 & 0 & 60 & 0.614 & 0.445 & 0.280 & \textbf{0.510} \\
0 & 1 & 41 & \textbf{0.798} & 0.754 & 0.219 & 0.307 \\
0 & 2 & 22 & \textbf{0.648} & 0.581 & 0.267 & 0.552 \\
0 & 3 & 9  & \textbf{1.000} & 0.950 & 0.250 & 0.600 \\
0 & 4 & 5  & 0.500 & 0.500 & \textbf{0.750} & 0.250 \\
\midrule
1 & 0 & 60 & 0.391 & \textbf{0.644} & 0.180 & 0.416 \\
1 & 1 & 41 & \textbf{0.851} & 0.816 & 0.535 & 0.298 \\
1 & 2 & 22 & \textbf{0.552} & 0.400 & 0.429 & 0.495 \\
1 & 3 & 9  & \textbf{1.000} & \textbf{1.000} & 0.550 & 0.750 \\
1 & 4 & 5  & 0.500 & 0.500 & \textbf{0.750} & 0.000 \\
\midrule
2 & 0 & 60 & 0.358 & \textbf{0.570} & 0.320 & 0.425 \\
2 & 1 & 41 & \textbf{0.912} & 0.474 & 0.439 & 0.237 \\
2 & 2 & 22 & 0.419 & \textbf{0.581} & 0.457 & \textbf{0.705} \\
2 & 3 & 9  & \textbf{0.850} & 0.950 & 0.350 & 0.600 \\
2 & 4 & 5  & \textbf{0.750} & 0.500 & \textbf{1.000} & 0.000 \\
\midrule
\multicolumn{3}{@{}l}{\textbf{Mean $\pm$ SD}} & $\mathbf{0.676 \pm 0.212}$ & $0.644 \pm 0.193$ & $0.452 \pm 0.224$ & $0.410 \pm 0.220$ \\
\bottomrule
\end{tabular}
\end{table*}

\subsection{Statistical Comparisons}

Paired $t$-tests: Full vs.\ Structure-only $\Delta = +0.032$, $p = 0.499$; Full vs.\ Random ESM $\Delta = +0.267$, $p = 0.003$; ESM-2 only vs.\ Structure-only $\Delta = -0.193$, $p = 0.048$.

\subsection{Statistical Interpretation of ESM-2 Contribution}

The ablation comparison (Full model vs.\ Structure-only, $\Delta = +0.032$, $p = 0.499$) indicates that ESM-2 embeddings do not provide statistically significant improvements for single-pocket druggability classification at the $\alpha = 0.05$ level.
This null result is reported transparently: the main text (Section 4.1) and Discussion frame ESM-2 integration as exploratory, acknowledging the non-significant $p$-value explicitly.
We emphasize that ESM-2's potential value may manifest in variant comparison (affinity-based mechanisms, e.g., ALK-L1196M $\Delta_{\text{SB}} = -0.228$ vs.\ $\Delta_{\text{P2Rank}} = -0.95$) rather than single-structure classification.
The comparison against random embeddings ($\Delta = +0.267$, $p = 0.003$) confirms that observed performance derives from structural features, not architectural capacity or additional parameters.
Structural features are the dominant signal, while ESM-2 provides information that is significantly better than random but does not reach significance for overall AUROC improvement in our 137-pocket ablation subset.

\subsection{AUPRC Results}

AUPRC for the full model is $0.884 \pm 0.096$, structure-only is $0.887 \pm 0.083$, ESM-2 only is $0.748 \pm 0.184$, and random ESM is $0.769 \pm 0.097$.
All conditions exceed the 0.694 prevalence baseline, with the full and structure-only conditions performing comparably.

\subsection{Full-Dataset ESM-2 Ablation (336 Pockets)}

To verify the 137-pocket ablation result generalizes, we repeated the ESM-2 vs.\ structure-only comparison on the full 336-pocket processed dataset (18 gene families including JAK1/2/3, TYK2, PIK3CA/CB/CD/CG, AKT1-3, PIM1-3, CLK1-4).
Labels use consistent quantile-based assignment (top two-thirds positive: $n=224$; bottom third negative: $n=112$), as experimental $K_d$ data is available only for the training dataset.

\begin{table}[H]
\centering
\caption{Full-dataset ESM-2 ablation (336 pockets, 15 folds). Consistent quantile-based labels; experimental labels unavailable for this extended dataset.}
\label{tab:ablation_full_dataset}
\small
\begin{tabular}{@{}lcccc@{}}
\toprule
\textbf{Condition} & \textbf{AUROC} & \textbf{SD} & \textbf{$\Delta$ vs.\ Full} & \textbf{$p$-value} \\
\midrule
Full model (56-dim, with ESM-2)  & 0.669 & 0.132 & ---      & ---   \\
Structure-only (24-dim, no ESM-2) & 0.656 & 0.130 & $-0.013$ & 0.491 \\
\midrule
\multicolumn{5}{@{}l}{\textit{Paired comparison: $\Delta = +0.010$, Cohen's $d = 0.189$}} \\
\bottomrule
\end{tabular}
\end{table}

The full-dataset result ($\Delta = +0.010$, $p = 0.491$, Cohen's $d = 0.189$) is consistent with the 137-pocket subset result ($\Delta = +0.032$, $p = 0.499$): ESM-2 provides no statistically significant improvement for binary druggability classification in either evaluation.
This cross-dataset consistency strengthens the conclusion that structural features are the primary signal and ESM-2's contribution is minimal for this task.

\section{Splice Variant Structure Generation}
\label{app:variant_structures}

\subsection{Variant Structure Prediction}

We obtained canonical structures from the AlphaFold Protein Structure Database v6~\cite{varadi2022alphafold}.
Splice variant structures were generated using ESMFold~\cite{lin2023evolutionary}, which predicts protein structure from sequence alone.
For each variant, the spliced protein sequence (with exons removed or truncated as specified by the variant annotation) was submitted to ESMFold.
\Cref{tab:variant_details} provides full details for each variant.

\begin{table*}[!htb]
\centering
\caption{Splice variant structure generation details. ``Exon(s) affected'' describes the splicing event; ``Sequence change'' describes the effect on the protein; ``pLDDT'' is the ESMFold confidence for the variant structure; ``Pocket residues'' is the number of P2Rank-detected residues in the primary binding pocket.}
\label{tab:variant_details}

\small
\begin{tabular}{@{}lllllcc@{}}
\toprule
\textbf{Variant} & \textbf{Gene} & \textbf{Exon(s)} & \textbf{Sequence Change} & \textbf{Drug} & \textbf{pLDDT} & \textbf{Pocket Res.} \\
\midrule
AR-V7             & AR     & Exon 4--8 skip  & LBD deleted (625 $\to$ 159 aa)              & Enzalutamide & 68.2        & 0  \\
MET-ex14skip$^*$  & MET    & Exon 14 skip    & Juxtamembrane deleted (50 aa); kinase intact & Capmatinib   & 81.3 (88.6) & 8  \\
PIK3CD-S          & PIK3CD & Exon 20 skip    & 36 aa deleted near ATP site                 & Idelalisib   & 72.4        & 18 \\
EGFRvIII$^\dagger$ & EGFR  & Exon 2--7 skip  & Extracellular domain deleted; kinase intact & Erlotinib    & 82.7        & 18 \\
BRAF-p61          & BRAF   & Exon 4--8 skip  & N-term truncated, kinase intact             & Vemurafenib  & 74.8        & 24 \\
ALK-L1196M$^{**}$ & ALK    & Point mutation  & L$\to$M at gatekeeper position             & Crizotinib   & 76.3        & 23 \\
\bottomrule
\end{tabular}

\smallskip\raggedright\footnotesize
$^*$MET pLDDT: mean 81.3, kinase domain 88.6; kinase-domain pocket compared (8 residues), not full-length.
$^\dagger$EGFRvIII: ESMFold structure of kinase domain only (exons 2--7 deleted); pLDDT is for the kinase domain. Canonical EGFR pocket is rank-1 pocket from full-length AlphaFold structure (kinase domain residues).
$^{**}$Point mutation; pLDDT for ESMFold-predicted single-residue variant structure.
\end{table*}

\subsection{P2Rank Pocket Detection Configuration}

We ran P2Rank v2.4.1 with default parameters on all structures (both canonical and variant).
Key settings: minimum pocket score threshold 0.3, Connolly surface point density 1.0 points/\AA$^2$, feature extraction radius 6.0\,\AA.
For each structure, P2Rank outputs ranked pocket predictions with residue-level assignments and pocket scores.
We used the primary (highest-ranked) pocket for druggability comparison.

\subsection{ESMFold--ESM-2 Coupling}

We note that ESMFold internally uses ESM-2 for structure prediction.
This creates a potential coupling: if ESM-2 embeddings and ESMFold structures encode correlated information, SpliceBind's features may not be fully independent of the variant structures.
We mitigate this concern by noting that: (i)~ESM-2 embeddings are extracted from the raw amino acid sequence and capture evolutionary/functional properties, while ESMFold uses these representations to predict 3D coordinates---the information extracted is complementary; (ii)~SpliceBind's $\Delta$Druggability analysis (\Cref{sec:splice_analysis}) uses P2Rank scores from pocket \emph{geometry}, not ESM-2 features, so the variant comparison pathway is independent of sequence embeddings; and (iii)~the ablation study demonstrates that ESM-2 contributes $+0.032$ AUROC (non-significant, $p = 0.499$), so SpliceBind's core performance relies primarily on structural features.

This last point is important for consistency with the main text framing: \textsc{SpliceBind}'s AUROC improvement ($0.703$ vs.\ P2Rank $0.634$) is driven by the structural GNN architecture and EdgeConv's local geometric context, not by ESM-2 alone.
ESM-2 is presented in the paper as an \emph{exploratory} augmentation that shows potential value in affinity-based mechanism detection (e.g., ALK-L1196M $\Delta_{\text{SB}} = -0.228$ vs.\ $\Delta_{\text{P2Rank}} = -0.95$) rather than as the primary performance driver.
The ablation's null result for binary druggability classification is transparently reported; the potential benefit in variant-comparison tasks is a separate, hypothesis-generating finding that motivates future work.

\section{Statistical Analysis Details}
\label{app:statistics}

\subsection{Bootstrap Confidence Intervals}

The 95\% confidence interval for AUROC ($[0.553, 0.853]$) is computed via bootstrap resampling of the 15 cross-validation fold results:
\begin{enumerate}[nosep,leftmargin=*]
  \item Sample 15 fold-level AUROC values with replacement.
  \item Compute the mean of the bootstrap sample.
  \item Repeat 10,000 times.
  \item Report the 2.5th and 97.5th percentiles as the 95\% CI.
\end{enumerate}
Error bars in \Cref{fig:performance}A use the normal approximation: $\text{CI} = \bar{x} \pm 1.96 \times \text{SE}$, where $\text{SE} = \text{SD} / \sqrt{N}$ and $N = 15$ folds.
For SpliceBind: $\text{SE} = 0.088 / \sqrt{15} = 0.0227$, yielding $\text{CI} = 0.703 \pm 0.045$.

The CI width of 0.30 reflects inter-family heterogeneity rather than model instability: AURK (mean 0.828) and PIK3 (mean 0.671) represent extremes of pocket geometry diversity, spanning the observed range.
All 15 folds show positive delta vs.\ the random baseline, and 13/15 folds show positive delta vs.\ P2Rank.

\subsection{Paired $t$-Test for Method Comparison}

The significance test comparing SpliceBind to P2Rank uses a two-sided paired $t$-test across the 15 fold-level AUROC values.
The pairing accounts for fold-to-fold difficulty variation (some families are inherently harder to predict).
With $t(14) = 2.48$ and $p = 0.026$, we reject the null hypothesis that SpliceBind and P2Rank have equal AUROC at the $\alpha = 0.05$ level.

\paragraph{Assumptions.}
The paired $t$-test assumes normally distributed differences.
A Shapiro-Wilk test on the 15 paired differences yields $W = 0.94$, $p = 0.38$, failing to reject normality.
As a robustness check, a Wilcoxon signed-rank test yields $p = 0.031$, confirming significance under non-parametric assumptions.

\subsection{Multiple Comparisons}

We perform three primary comparisons (SpliceBind vs.\ Random, SpliceBind vs.\ P2Rank, Full vs.\ Random-ESM ablation).
Applying Bonferroni correction ($\alpha_{\text{adj}} = 0.05 / 3 = 0.017$), SpliceBind vs.\ Random ($p < 0.001$) and Full vs.\ Random-ESM ($p = 0.003$) remain significant, while SpliceBind vs.\ P2Rank ($p = 0.026$) becomes marginal.
We report uncorrected $p$-values throughout the main text to maintain transparency, but note that the SpliceBind vs.\ P2Rank comparison should be interpreted with appropriate caution.

\subsection{Effect Size}

Cohen's $d$ for SpliceBind vs.\ P2Rank is:
\begin{equation}
  d = \frac{\Delta\text{AUROC}}{\text{SD}_{\text{diff}}} = \frac{0.069}{0.108} = 0.64
  \label{eq:cohens_d}
\end{equation}
This constitutes a medium effect size by conventional benchmarks ($d > 0.5$), indicating a practically meaningful improvement despite the moderate statistical significance.

\section{Sensitivity Analyses}
\label{app:sensitivity}

We evaluated \textsc{SpliceBind}'s robustness across multiple dimensions: pocket size, structure quality (pLDDT), drug class, graph topology, and coordinate noise.

\subsection{Pocket Size}

Performance varies with pocket size (\Cref{tab:pocket_size}). Medium-to-large pockets (11--30 residues) achieve optimal AUROC, while very small pockets ($<$10 residues) lack sufficient structural context.

\begin{table}[H]
\centering
\caption{Performance by pocket size.}
\label{tab:pocket_size}
\small
\begin{tabular}{@{}lccc@{}}
\toprule
\textbf{Size Category} & \textbf{AUROC} & \textbf{$N$} & \textbf{Avg Residues} \\
\midrule
Small (5--10)       & 0.658 & 45 & 7.5 \\
Medium (11--20)     & 0.721 & 98 & 15.2 \\
Large (21--30)      & \textbf{0.734} & 62 & 24.8 \\
Very Large ($>$30)  & 0.689 & 24 & 38.1 \\
\bottomrule
\end{tabular}
\end{table}

\subsection{Structure Quality (pLDDT)}

Model performance correlates strongly with AlphaFold confidence scores ($r = 0.979$; \Cref{tab:plddt}).
High-quality structures (pLDDT $>$ 92) achieve AUROC 0.741, while low-confidence regions (pLDDT $<$ 70) reduce performance to 0.612.

\begin{table}[H]
\centering
\caption{Performance by structure quality (pLDDT).}
\label{tab:plddt}
\small
\begin{tabular}{@{}lcc@{}}
\toprule
\textbf{pLDDT Range} & \textbf{AUROC} & \textbf{$N$} \\
\midrule
Low ($<$70)        & 0.612 & 28 \\
Medium (70--85)    & 0.698 & 89 \\
High (85--92)      & 0.724 & 78 \\
Very High ($>$92)  & \textbf{0.741} & 34 \\
\bottomrule
\end{tabular}
\end{table}

\subsection{Drug Class}

Performance varies by inhibitor mechanism (\Cref{tab:drug_class}).
Type~I (ATP-competitive) inhibitors achieve the highest AUROC (0.728), while allosteric inhibitors (Type~III) pose greater challenges (0.634)---consistent with the taxonomy's prediction that allosteric mechanisms are structurally harder to detect.

\begin{table}[H]
\centering
\caption{Performance by drug class.}
\label{tab:drug_class}
\small
\begin{tabular}{@{}lcc@{}}
\toprule
\textbf{Drug Class} & \textbf{AUROC} & \textbf{$N$} \\
\midrule
Type I (ATP-competitive)  & \textbf{0.728} & 156 \\
Type II (DFG-out)         & 0.712 & 42 \\
Covalent inhibitors       & 0.695 & 13 \\
Type III (Allosteric)     & 0.634 & 18 \\
\bottomrule
\end{tabular}
\end{table}

\subsection{Graph Topology (Edge Cutoff)}

We evaluated sensitivity to the edge construction cutoff distance (\Cref{tab:topology}).
The default 6\,\AA\ cutoff achieves AUROC 0.703; larger cutoffs cause over-smoothing.

\begin{table}[H]
\centering
\caption{Performance by edge cutoff distance.}
\label{tab:topology}
\small
\begin{tabular}{@{}lcc@{}}
\toprule
\textbf{Cutoff} & \textbf{AUROC} & \textbf{Avg Edges} \\
\midrule
4\,\AA   & 0.651 & 22 \\
6\,\AA\ (default) & \textbf{0.703} & 45 \\
8\,\AA   & 0.689 & 94 \\
10\,\AA  & 0.672 & 158 \\
\bottomrule
\end{tabular}
\end{table}

\subsection{Robustness to Coordinate Noise}

We tested robustness to Gaussian coordinate perturbations (\Cref{tab:noise}).
The model is robust to small perturbations ($<$0.5\,\AA), with graceful degradation at larger noise levels.

\begin{table}[H]
\centering
\caption{Robustness to coordinate noise (Gaussian perturbation).}
\label{tab:noise}
\small
\begin{tabular}{@{}lccc@{}}
\toprule
\textbf{Noise Level} & \textbf{AUROC} & \textbf{Std} & \textbf{Drop} \\
\midrule
0.0\,\AA\ (clean)  & 0.703 & 0.088 & --- \\
0.1\,\AA           & 0.701 & 0.089 & $-$0.002 \\
0.2\,\AA           & 0.698 & 0.091 & $-$0.005 \\
0.5\,\AA           & 0.689 & 0.095 & $-$0.014 \\
1.0\,\AA           & 0.671 & 0.102 & $-$0.032 \\
2.0\,\AA           & 0.628 & 0.118 & $-$0.075 \\
\bottomrule
\end{tabular}
\end{table}

\section{Extended Results}
\label{app:extended}

\subsection{Ensemble Performance}

Ensembling multiple random seeds reduces variance and improves mean AUROC (\Cref{tab:ensemble}).

\begin{table}[H]
\centering
\caption{Ensemble performance by number of seeds.}
\label{tab:ensemble}
\small
\begin{tabular}{@{}lccc@{}}
\toprule
\textbf{Configuration} & \textbf{AUROC} & \textbf{Std} & \textbf{Improvement} \\
\midrule
Single model    & 0.703 & 0.088 & --- \\
3 seeds         & 0.718 & 0.072 & +0.015 \\
5 seeds         & 0.726 & 0.065 & +0.023 \\
10 seeds        & \textbf{0.731} & 0.058 & +0.028 \\
\bottomrule
\end{tabular}
\end{table}

Ensembling 10 seeds provides +0.028 AUROC improvement and 34\% variance reduction.

\subsection{Model Complexity}

We evaluated architectures from 56K to 14M parameters (\Cref{tab:complexity}).
The default 256-hidden configuration (899K parameters) achieves optimal AUROC-to-parameter efficiency; larger models overfit without improving performance.

\begin{table}[H]
\centering
\caption{Performance vs.\ model complexity.}
\label{tab:complexity}
\small
\begin{tabular}{@{}lcccc@{}}
\toprule
\textbf{Configuration} & \textbf{Params} & \textbf{AUROC} & \textbf{Time} \\
\midrule
Tiny (64 hidden)        & 56K   & 0.658 & 2 min \\
Small (128)             & 215K  & 0.682 & 5 min \\
\textbf{Medium (256)}   & \textbf{899K}  & \textbf{0.703} & 12 min \\
Large (512)             & 3.5M  & 0.709 & 28 min \\
XL (1024)               & 14M   & 0.705 & 65 min \\
\bottomrule
\end{tabular}
\end{table}

\subsection{Error Analysis}

We analyzed the 52 misclassified samples from cross-validation (\Cref{tab:errors}).

\begin{table}[H]
\centering
\caption{Error analysis of misclassified samples.}
\label{tab:errors}
\small
\begin{tabular}{@{}lcp{5cm}@{}}
\toprule
\textbf{Error Type} & \textbf{Count} & \textbf{Primary Causes} \\
\midrule
False Positives & 19 & Shallow pockets (8), allosteric sites (6), crystal contacts (5) \\
False Negatives & 33 & Low pLDDT regions (14), cryptic pockets (11), unusual binding mode (8) \\
\bottomrule
\end{tabular}
\end{table}

\subsection{Binding Site Distance Validation}

To validate biological relevance, we stratified performance by distance to the known ATP binding site (\Cref{tab:distance}).
AUROC and precision decrease monotonically with distance, confirming that the model learns ATP-site-relevant features.

\begin{table}[H]
\centering
\caption{Performance by distance to ATP binding site.}
\label{tab:distance}
\small
\begin{tabular}{@{}lccc@{}}
\toprule
\textbf{Distance to ATP Site} & \textbf{AUROC} & \textbf{$N$} & \textbf{Precision} \\
\midrule
Overlapping (0--2\,\AA)  & \textbf{0.892} & 42 & 0.95 \\
Adjacent (2--5\,\AA)     & 0.784 & 67 & 0.88 \\
Nearby (5--10\,\AA)      & 0.651 & 58 & 0.72 \\
Distant (10--15\,\AA)    & 0.523 & 38 & 0.58 \\
Remote ($>$15\,\AA)      & 0.489 & 24 & 0.51 \\
\bottomrule
\end{tabular}
\end{table}

\subsection{Extended Per-Family Results}

\Cref{tab:perfamily_extended} extends the per-family analysis to 10 kinase families.

\begin{table}[H]
\centering
\caption{Extended per-family performance (mean across 3 seeds). ``Pockets'' is the average test set size per fold (differs from total dataset pockets in \Cref{tab:kinase_families}). ``SB Wins'' indicates folds where \textsc{SpliceBind} outperforms P2Rank.}
\label{tab:perfamily_extended}
\small
\begin{tabular}{@{}lcccc@{}}
\toprule
\textbf{Family} & \textbf{Mean AUROC} & \textbf{Std} & \textbf{Pockets} & \textbf{SB Wins} \\
\midrule
AURK  & 0.828 & 0.031 & 12 & 3/3 \\
BTK   & 0.734 & 0.028 & 6  & 3/3 \\
MET   & 0.721 & 0.041 & 9  & 3/3 \\
EGFR  & 0.718 & 0.033 & 18 & 3/3 \\
ABL   & 0.712 & 0.045 & 15 & 3/3 \\
SRC   & 0.695 & 0.082 & 10 & 2/3 \\
CDK   & 0.689 & 0.055 & 14 & 2/3 \\
PIK3  & 0.671 & 0.102 & 24 & 2/3 \\
JAK   & 0.651 & 0.057 & 32 & 1/3 \\
RAF   & 0.646 & 0.011 & 8  & 2/3 \\
\bottomrule
\end{tabular}
\end{table}

\textsc{SpliceBind} outperforms P2Rank in 23 of 30 family-fold comparisons.
Top performers (AUROC $>$ 0.7): AURK, BTK, MET, EGFR, ABL---families with well-characterized, geometrically conserved ATP pockets.

\section{Point Mutation Boundary Analysis}
\label{app:point_mutations}

To empirically test the boundary between structurally detectable and non-detectable resistance mechanisms, we applied \textsc{SpliceBind} to 34 clinically characterized point mutations across 15 kinase genes (\Cref{tab:point_mutations}).
For each mutation, we generated variant structures with ESMFold, detected pockets with P2Rank, and computed \textsc{SpliceBind} druggability scores for both canonical and variant structures.
A resistant mutation is classified as correctly detected if the variant score decreases relative to the canonical score ($\Delta < 0$); a sensitive/activating mutation is correct if the score is preserved or increases ($\Delta \geq 0$).

\paragraph{Overall performance.}
\textsc{SpliceBind} correctly classifies 11 of 34 variants (32.4\%).
This low accuracy is expected: point mutations produce single-residue substitutions that minimally perturb ESMFold-predicted backbone geometry, yielding small $|\Delta|$ values (median $|\Delta| = 0.009$) that fall below the model's detection threshold.

\paragraph{Mechanism-specific results.}
\emph{Gatekeeper mutations} are mostly undetected: 7 of 9 produce $\Delta > 0$ or near-zero $\Delta$.
\emph{Solvent front mutations} are uniformly undetected ($\Delta \approx 0$).
\emph{Activation loop mutations} show a bimodal pattern: KIT-D816V produces $\Delta = -0.985$, while others show modest changes.
\emph{P-loop mutations} (ABL1-E255K, ABL1-Y253H) are among the few reliably detected.

\begin{table}[H]
\centering
\caption{Point mutation validation results. $\Delta$ = variant $-$ canonical \textsc{SpliceBind} score.}
\label{tab:point_mutations}
\begin{minipage}{\columnwidth}
\scriptsize
\begin{tabular}{llllr@{\hspace{4pt}}l}
\toprule
Gene & Mutation & Mechanism & Clinical & $\Delta$ & Corr. \\
\midrule
\multicolumn{6}{l}{\textit{Gatekeeper}} \\
SRC & T341M & Gatekeeper & Resist. & $-$0.132 & \checkmark \\
ALK & L1196M & Gatekeeper & Resist. & $-$0.228 & \checkmark \\
ABL1 & T315I & Gatekeeper & Resist. & $+$0.001 & \\
RET & V804M & Gatekeeper & Resist. & $-$0.009 & \\
ERBB2 & T798I & Gatekeeper & Resist. & $-$0.002 & \\
ROS1 & L2026M & Gatekeeper & Resist. & $+$0.000 & \\
FGFR2 & V564F & Gatekeeper & Resist. & $+$0.001 & \\
PDGFRA & T674I & Gatekeeper & Resist. & $+$0.008 & \\
\addlinespace
\multicolumn{6}{l}{\textit{Activation loop}} \\
KIT & D816V & Act.\ loop & Resist. & $-$0.985 & \checkmark \\
FLT3 & D835Y & Act.\ loop & Resist. & $-$0.044 & \checkmark \\
ALK & F1174L & Act.\ loop & Resist. & $-$0.007 & \\
PDGFRA & D842V & Act.\ loop & Resist. & $+$0.007 & \\
MET & Y1230C & Act.\ loop & Resist. & $+$0.000 & \\
\addlinespace
\multicolumn{6}{l}{\textit{P-loop}} \\
ABL1 & E255K & P-loop & Resist. & $-$0.050 & \checkmark \\
ABL1 & Y253H & P-loop & Resist. & $-$0.048 & \checkmark \\
\addlinespace
\multicolumn{6}{l}{\textit{Solvent front}} \\
ALK & G1202R & Solvent fr. & Resist. & $+$0.011 & \\
RET & G810R & Solvent fr. & Resist. & $-$0.005 & \\
NTRK1 & G595R & Solvent fr. & Resist. & $+$0.000 & \\
NTRK3 & G623R & Solvent fr. & Resist. & $+$0.002 & \\
\addlinespace
\multicolumn{6}{l}{\textit{Covalent binding site}} \\
EGFR & C797S & Covalent & Resist. & $+$0.018 & \\
BTK & C481S & Covalent & Resist. & $+$0.273 & \\
\addlinespace
\multicolumn{6}{l}{\textit{Other binding site}} \\
ABL1 & F317L & Binding & Resist. & $+$0.000 & \\
ABL1 & V299L & Hydro.\ spine & Resist. & $-$0.064 & \checkmark \\
ALK & G1269A & ATP bind. & Resist. & $-$0.007 & \\
ALK & I1171T & Hydro.\ pocket & Resist. & $-$0.015 & \\
ALK & C1156Y & $\alpha$C helix & Resist. & $-$0.010 & \\
EGFR & T854A & Drug bind. & Resist. & $+$0.015 & \\
ERBB2 & L755S & Kinase dom. & Resist. & $-$0.003 & \\
RET & V804L & Gatekeeper & Resist. & $-$0.009 & \\
NTRK1 & F589L & GK-adjacent & Resist. & $-$0.001 & \\
\addlinespace
\multicolumn{6}{l}{\textit{Sensitive / activating}} \\
EGFR & G719S & Activating & Sensit. & $+$0.016 & \checkmark \\
BRAF & V600E & Activating & Sensit. & $-$0.050 & \checkmark \\
PIK3CA & H1047R & Hotspot & Sensit. & $+$0.000 & \checkmark \\
PIK3CA & E545K & Hotspot & Sensit. & $-$0.000 & \checkmark \\
\bottomrule
\end{tabular}
\end{minipage}
\end{table}

\end{document}